\documentclass[]{spie}  
\usepackage[]{graphicx}

\usepackage{color}
\usepackage{bm}

\usepackage[english]{babel}
\usepackage{amsmath}
\usepackage{amssymb}
\usepackage{units}
\usepackage{upgreek}
\usepackage[colorlinks=true, allcolors=blue]{hyperref}
\usepackage{ctable}
\usepackage{footnote}
\usepackage{bbold}
\usepackage{textcomp}
\usepackage[absolute,overlay]{textpos}
\usepackage{helvet}
\usepackage{svg}
\newcommand{\orcid}[1]{\href{https://orcid.org/#1}{\includesvg[width=11pt]{orcid}}}

\title{Optical tools for laser machining along six orders of magnitude} 

\author{Julian Hellstern,\supit{a} Christoph Tillkorn,\supit{a}\, Tim Hieronymus,\supit{b}\ Myriam Kaiser,\supit{b} Torsten Beck,\supit{b} and Daniel Flamm\supit{b} 
\skiplinehalf
\supit{a}TRUMPF Laser GmbH, Aichhalder Str.\,39, 78713 Schramberg, Germany\\
\supit{b}TRUMPF Laser- und Systemtechnik GmbH, Johann-Maus-Str.\,2, 71254 Ditzingen, Germany
}

\authorinfo{Further author information:\\ E-Mail: \href{mailto:Julian.Hellstern@trumpf.com}{julian.hellstern@trumpf.com}}

\begin{document} 
  \maketitle 

\begin{abstract}
We present an overview on the development and characterization of multiscale laser processing optics for versatile material modifications across more than six orders of magnitude. Starting with solutions for micromachining we present high-NA microscope objectives creating sub-wavelength material modifications on macroscopic scales with highest peak intensities. Moving on to the millimeter range, the adaptability and scalability of scanning optics is examined for large-area machining. Finally, we explore line beam optics in the meter range, evaluating their use in uniform material processing using average powers above $\unit[100]{kW}$. This study provides an insight into the design and performance characteristics of such optics and demonstrates their potential in advanced laser processing.
\end{abstract}


\keywords{Beam shaping, ultrafast optics, laser materials processing, scanning heads, surface treatment, line beam optics.}

\section{INTRODUCTION}

\label{sec:intro}  
\begin{textblock*}{16cm}(2.67cm,1,7cm) 
   \centering
  \small \textsf{Invited Paper}
\end{textblock*}
\begin{textblock*}{16cm}(2.67cm,1cm) 
   \centering
  \tiny \textsf{Julian Hellstern, Christoph Tillkorn, Tim Hieronymus, Myriam Kaiser, Torsten Beck, and Daniel Flamm ``Optical tools for laser machining along six orders of magnitude'', Proc. SPIE 12878, High-Power Laser Materials Processing: Applications, Diagnostics, and Systems XIII, 1287802 (12 March 2024); \url{https://doi.org/10.1117/12.3002211}.}
\end{textblock*}

\begin{textblock*}{17cm}(2.25cm,25.25cm) 
   \centering \small 
   \textsf{
   © 2024 Society of Photo‑Optical Instrumentation Engineers (SPIE). One print or electronic copy may be made for personal use only. Systematic reproduction and distribution, duplication of any material in this publication for a fee or for commercial purposes, and modification of the contents of the publication are prohibited. \\
   High-Power Laser Materials Processing: Applications, Diagnostics, and Systems XIII © 2024 SPIE. \url{https://doi.org/10.1117/12.3002211}.}
\end{textblock*}
\noindent
Multi-kilowatt continuous-wave (cw) lasers have undergone a remarkable commoditization process over the past decade. This is clearly seen from dollar-per-watt costs of output power dropping one order of magnitude within the last decade.\cite{2023ThossLaserMarket} Considering, e.g.~fiber laser systems, costs have already fallen well below the $10$-dollar-per-watt limit.\cite{2023ThossLaserMarket} The $\unit[100]{kW}$-average power barrier was broken in 2013 at the latest.\cite{shcherbakov2013industrial} Useful non-military applications, however, are rather rare and may be found in welding of sheet steel in extreme geometries\cite{kawahito2018ultra}.
\par
A continuous increase in performance and a simultaneous reduction in costs will also affect other laser architectures. For example, there is constant news about extreme power and energy records for ultrashort laser systems \cite{saraceno2019amazing, stark20211, dominik2022thin}. Even considering industrial availability, in the near future, $\sim\unit[100]{mJ}$-class lasers operating in the multi-kilowatt-regime providing subpicosecond pulses will be available \cite{dominik2022thin}. Here, similar to the cw lasers mentioned above, radiation has reached such extreme levels that useful applications will always require additional tools to trigger a specific light-matter-interaction.\cite{flamm2021structured, Wang2024PR} One main enabler is the processing optics allowing to distribute the radiation onto large surfaces or into large volumes.\cite{kumkar2017throughput} In this work, a thorough review is provided about a variety of industry-grade processing heads for laser machining along six spatial orders of magnitude.
\par
Starting at the micrometer range (Sec.~\ref{sec:micro}), we present optics designed for ultrashort pulse applications providing peak intensities in the order of $\unit[1\text{E}14]{W/cm^2}$ at $\sim\unit[1]{ps}$. These optics are, for example, capable of performing in-volume modifications of transparent materials, making them suitable for display applications.\cite{flamm2022protecting} To achieve modifications at the micrometer scale while maintaining large working volumes, high-NA focusing units are employed.\cite{Flamm2022AOT}
\par 
Covering the millimeter range (Sec.~\ref{sec:milli}), we explore the versatility and scalability of scanning optics for remote welding applications.\cite{moller2022novel} The required F-Theta lenses focus cw radiation up to $\unit[20]{kW}$ in average power. The processing head can be equipped with beam shaping capabilities, such as multiple spots\cite{kumkar2017throughput,moller2022novel,moller2022spatially}, ring distributions\cite{grunewald2021influence}, and 2-in-1 fibers\cite{bocksrocker2019reduction} to satisfy the versatility of applications.
\par 
Finally, we discuss optical solutions for processing in the meter range for applications like thermal annealing\cite{IRLinie} or laser-induced forward transfer (Sec.~\ref{sec:meter}).\cite{serra2019laser} Here, highest uniformities along the meter-scaled line focus require a well-balanced design incorporating concepts such as coherence management and beam homogenization.\cite{10.1117/12.2608238} Line focus concepts are well suited for processing in the pulsed (nanoseconds) and cw regime. For the latter case, we report on thermal annealing with $\unit[144]{kW}$ in average power from a multi-laser system resulting in line densities above $\unit[40]{W/mm}$.\cite{IRLinie}
\par
In any of the three following main sections we first provide insights into the focus unit, discuss corresponding beam shaping concepts and, finally, present selected laser machining strategies. 

\section{Modifications in the micrometer scale}\label{sec:micro}
The term micromachining covers laser-induced material modifications in the range of less than one millimeter typically generated from short- or ultrashort laser pulses.\cite{gattass2008femtosecond, itoh2006ultrafast, flamm2021structured} In the following, we will discuss optical tools that deal with the lower range of this spatial scale, i.e., in the single-digit micrometer range or even slightly below one micrometer. We will start introducing the focusing unit providing numerical apertures (NAs) up tp $\sim 0.4$ in Sec.~\ref{sec:MO} and will explain why they are the key to realizing different structured light concepts, see Sec.~\ref{sec:structuredmicro}. Finally, we will apply these concepts to radiation from ultrafast lasers and demonstrate advanced transparent materials processing, see Sec.~\ref{sec:glassproc}. 

\subsection{Large-working-volume microscope objective}\label{sec:MO}
When designing focusing units for micromachining with ultrashort laser pulses several side conditions must be taken into account. The primary focus is of course on generating highest peak intensities---in some cases even higher than $\unit[1\text{E}14]{W/cm^2}$.\cite{jenne2018high} In addition to resulting highest demands on materials and coatings, large working distances and---for the advanced beam shaping concepts, cf.~Sec.~\ref{sec:structuredmicro}---large working volumes are required. With regard to the latter point, micrometer-scale material modifications are required to be induced in a millimeter-scale volume. Thus, within this working volume $V_{\textbf{W}} \sim \unit[1]{mm^3}$ a diffraction-limited performance is a fundamental prerequisite for the microscope objective (MO) design. In practice, however, a large axial and lateral working volume lead to an inherent physical limitation and conflict at high numerical apertures. An absence of lateral aberrations requires the system to satisfy the Abbe sine condition.\cite{braat1997abbe} Hence residual spherical aberrations will occur over the axial range. On the other hand, the absence of axial aberrations requires the Herschel condition to be satisfied resulting in remaining lateral aberrations mainly composed by coma.\cite{braat1997abbe} In particular, the rotational symmetry of the individual foci is important for a controlled laser modification process, cf.~Sec.~\ref{sec:glassproc}. Therefore, the microscope lenses have been designed to satisfy the sine condition with minor residual spherical aberrations along the axial range.
\par 
In general, large working volumes will result in a complex mixture of field-dependent aberrations, such as coma, astigmatism, field curvature and distortion.\cite{gross2005handbook} Minimization of aberrations can be achieved when distributing the refractive power over multiple lenses reducing in particular spherical aberrations. Further design concepts include lens bending where large angle-of-incidences on lens surfaces are avoided by employing meniscus shapes, see MO examples in Sec.~\ref{sec:structuredmicro}. 
\par 
This section's focusing units are intended to be used for processing with ultrashort laser pulses. However, a chromatic compensation of the corresponding broad spectrum is not necessary as band-limited pulses with durations of $>\unit[300]{fs}$ are employed. This assumption also allows the use of a single type of glass---typically a UV-grade fused silica with damage thresholds above $ \unit[30]{J/cm^2}$ for picosecond pulses.\cite{jedamzik55122bulk}

\subsection{Structured light for micromachining}\label{sec:structuredmicro}
Diffraction limited spots in the single-digit micrometer scale result in Rayleigh lengths of $z_{\text{R}} \sim \unit[10]{\upmu m}$ at wavelengths $\lambda \sim \unit[1]{\upmu m}$. The associated processing strategies pose challenges for the machine integrator, as the focus position typically has to be maintained better than $z_{\text{R}}$. The optical solution for increased tolerance in the focus position is the use of so-called non-diffracting beams.\cite{mcgloin2005bessel, woerdemann2012structured} These are particularly important for machining transparent materials, as extreme aspect ratios enable single-pass processing of glass substrates.\cite{tsai2014internal, jenne2018high, rave2021glass, jenne2020facilitated} The generation of non-diffracting beams is typically axicon-based. When illuminated with coherent radiation a special interference pattern is generated directly behind the axicon tip.\cite{mcgloin2005bessel} Depending on the spatial dimensions of the illumination, aspect ratios might easily exceed $1000 : 1$ (length dimensions : width of the central on-axis maximum).\cite{flamm2019beam} However, using the non-diffracting beam directly from the axicon results in various processing disadvantages, such as, e.g., a vanishing working distance. For this reasons axicons are typically embedded into telescopic setups where the macroscopic non-diffracting beam is imaged with demagnification into or onto the workpiece. Here, too, large-working-volume MOs are used as they allow machining with working distances that equal the employed focal lengths, cf.~Sec.~\ref{sec:MO}.
\begin{figure}[]
    \centering
    \includegraphics[width=0.88\textwidth]{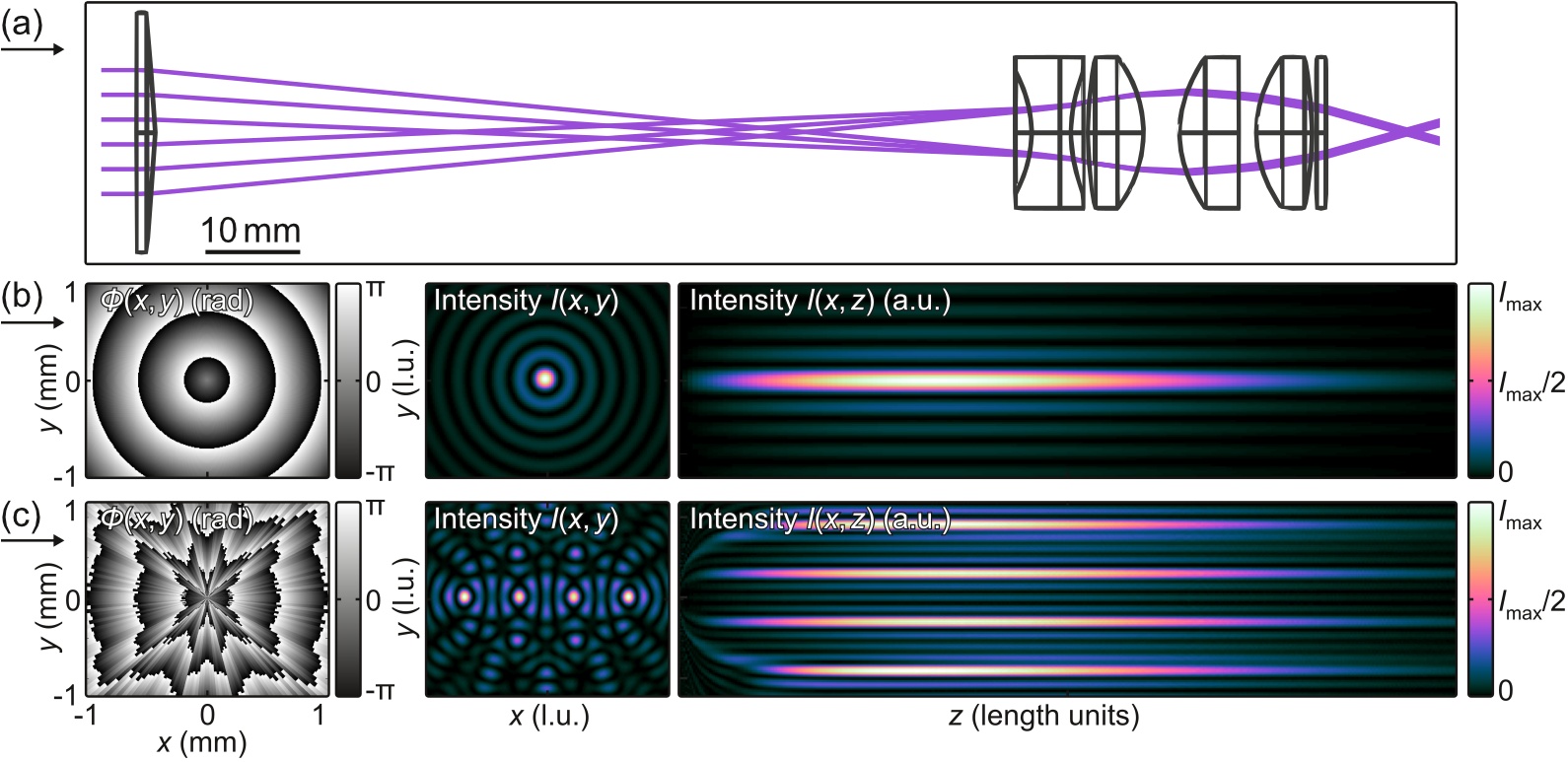}
    \caption{Ray optical representation of shaping non-diffracting beams (a). The axicon-like element is generating the far-field ring in the back-focal plane of an NA-0.4 microscope objective generating a millimeter-scaled non-diffracting beam in the focus. Bessel-Gaussian beam generation using a diffractive axicon\cite{leach2006generation, golub2006fresnel} (b) with phase-only transmission $\Phi\left(x,y\right)$, transverse $I\left(x,y\right)$ and longitudinal $I\left(x,z\right)$ intensity distribution.   
    Generalized axicon (c) for generating non-diffracting beams with tailored transverse intensity distributions. \cite{jenne2020facilitated, bergner2018time}}
    \label{fig:TC_gen}
\end{figure}
\par 
Figure \ref{fig:TC_gen} depicts the optical design of an optical head for non-diffracting beam processing. The axicon-like element (top) fulfills two functions. Firstly, it generates the non-diffracting beam and secondly, it performs its own far-field transformation. The combination of these two functionalities in one element has clear alignment advantages, as the optical axes are already aligned with each other.\cite{flamm2019beam} The far-field ring in the back-focal plane of the microscope objective is the prerequisite for the high-intense non-diffracting beam in the focus. Here, due to the objective's numerical aperture of $\sim 0.4$ we expect a transverse resolution of $d_0 \approx \unit[2]{\upmu m}$. For example, if the standard Bessel-Gaussian beam is generated from a Gaussian raw beam and a radial symmetric axicon, see Fig.~\ref{fig:TC_gen}\,(b), the width of the central on-axis maximum corresponds to $d_0$, too. Considering its length, it is well known that with the dimensions of the raw beam the length of the Bessel-Gaussian beam can be scaled linearly. Here, we expect to robustly shape a length spectrum from $l_0 = \unit[\left(0.2 \dots 2\right)]{mm}$.\cite{mcgloin2005bessel} By replacing the NA-0.4-MO ($f = \unit[10]{mm}$) with an NA-0.2-MO ($f = \unit[20]{mm}$) the maximum length can be further increased to $>\unit[8]{mm}$.\cite{flamm2019beam} \par 
The example of generating the Bessel-Gaussian beam as the most prominent non-diffracting beam already shows the importance designing the beam shaping element together with the focusing unit. This becomes even clearer if a more customized focus distribution is to be generated. Recently, we have introduced the concept of generalized axicons\cite{chen2019generalized, chen2022customized} where non-diffracting beams can be generated with process-adapted transverse intensity distributions. Here, all outstanding propagation properties of non-diffracting beams, such as extreme aspect ratio or self-healing\cite{mcgloin2005bessel} are maintained and, at the same time, the concept allows to tailor the transverse intensity profile $I\left(x,y\right)$. One example is depicted in the wave-optical representation in Fig.~\ref{fig:TC_gen}\,(c). Here, instead of one single on-axis intensity maximum (as for the Bessel-Gaussian case in (b)) four maxima of equal weights have been shaped at $y=0$. The generalized axicon causing this interference effect is also plotted as phase-only transmission with phase modulation $\Phi\left(x, y\right)$. At first glance, it is certainly not clear that an axicon is at hand. However, the phase mask shows the constant slope in the radial direction. The azimuthal dependency, however, is clearly altered in comparison to the radial symmetric case.\cite{chen2019generalized, flamm2020generalized} Generalized axicons, as the one depicted in Fig.~\ref{fig:TC_gen}\,(c), are preferebly generated diffractive optically or via digital holography.\cite{flamm2021structured}.
\par 
The imaging concept for shaping non-diffracting beams,  with long working distances, cf.~Fig.~\ref{fig:TC_gen}\,(a) represents only one useful case for the MO as part of a telescope within a laser processing head. More common is the direct focusing of a raw beam with the MO in a $2f$-like configuration acting as Fourier operator.\cite{flamm2021structured, flamm2022protecting, goodman2005introduction} Associated beam shaping concepts relate to homogenization, especially flat-top beam shaping,\cite{laskin2014beam, dickey2018laser} and beam splitting\cite{golub2004laser, flamm2022multi} or combinations thereof.\cite{flamm2021structured} Maximum flexibility is provided when 3D beam splitters are used allowing to freely distribute the laser energy in the volume of a focusing unit. Here in particular, a sophisticated optical concept is required ensuring diffraction limited spots which, in turn, generate comparable machining results for each of the focus copies.
\begin{figure}[]
    \centering
    \includegraphics[width=0.95\textwidth]{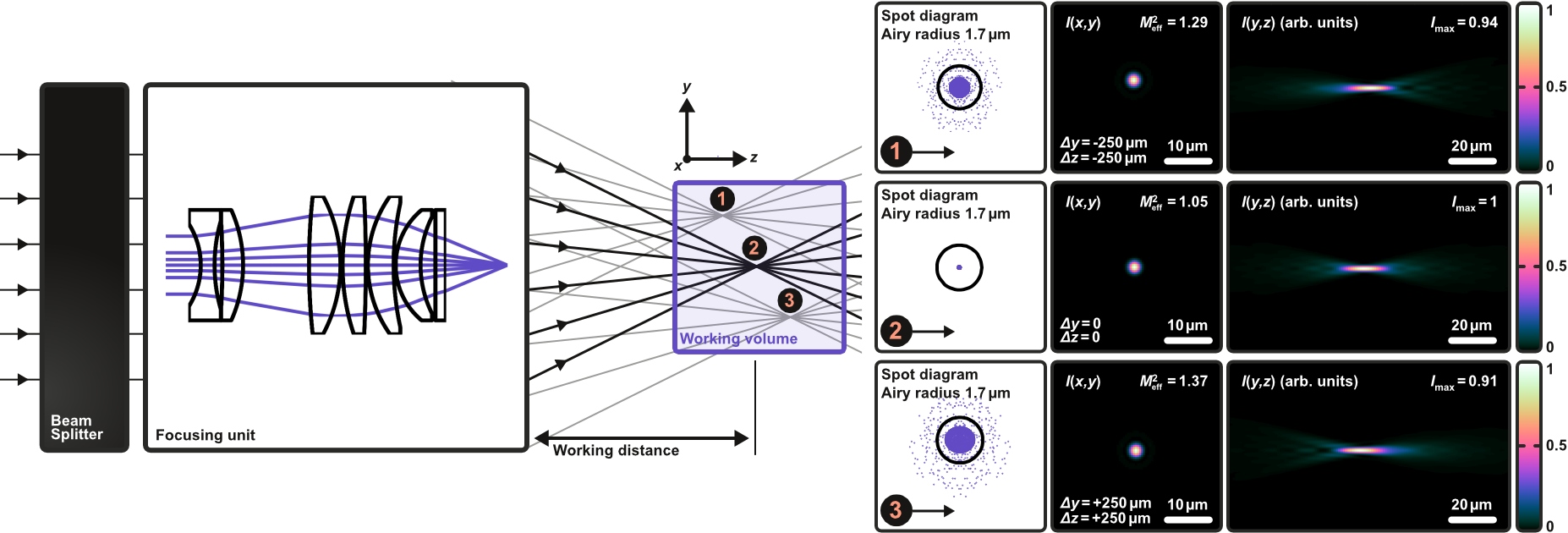}
    \caption{Focusing performance of the large-working-volume and large-working distance microscope objective (MO) exhibiting an effective focal length $f_{\text{eff}}=\unit[10]{mm}$ and $\text{NA}=0.45$. At three focus positions within a working volume of $V_{\text{W}} = \left(\unit[0.5]{mm}\right)^3$ similar spot qualities, peak intensities and symmetries are expected.}
    \label{fig:MOperf}
\end{figure}
Figure \ref{fig:MOperf} shows the large-working-volume MO implemented into a $2f$-like optical setup with the beam splitter in the MO's back focal plane. The focusing perfomance is investigated by considering ray- and wave-optically three focus positions within a volume of $V_{\text{W}} = d^3 = \left(\unit[0.5]{mm}\right)^3$. The focus at $\left(y=0, z=0\right)$ is on the geometrical focus of the MO serving as reference. For the sake of simplicity, beam splitting is carried out only in two dimensions ($y$- and $z$-axis) and $x=0$ is set. In any of the three considered cases, peak intensities $I_{\text{peak}}$, beam qualities $M^2_{\text{eff}}$\cite{ISO11146} and focus symmetries are very similar, see right hand side of Fig.~\ref{fig:MOperf}. We therefore expect that machining within the focal volume will result in similar processing results---a precondition for parallel processing and throughput scaling with multiple spots.\cite{flamm2021structured, flamm2022multi, Flamm2022AOT}
\par
\begin{figure}[]
    \centering
    \includegraphics[width=1.0\textwidth]{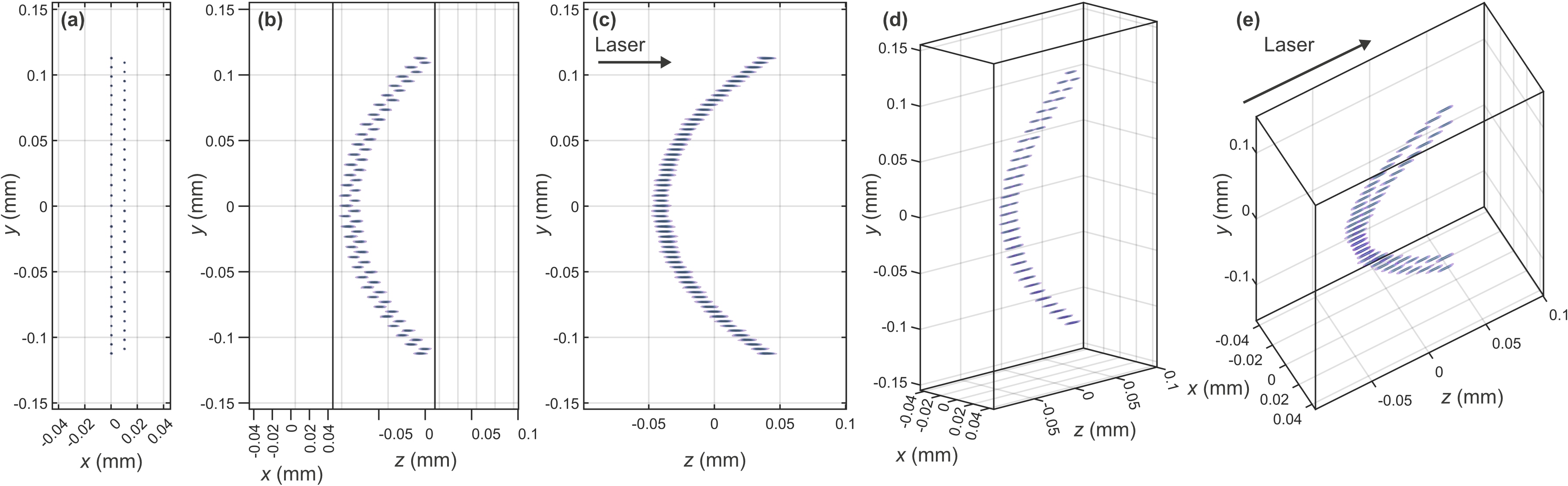}
    \caption{Iso-intensity representation $I\left(x,y,z\right)$ of a focus distribution composed of $60$ Gaussian foci distributed along a circular arc. Laser propagates parallel to the $z$-axis. The subfigures (a)\,--\,(e) show the same focus shape from different perspectives. }
    \label{fig:3D}
\end{figure}

 
One example demonstrating the flexibility of the 3D-beam splitting concept can be seen in Fig.~\ref{fig:3D} where an iso-intensity distribution shows one focus distribution from five different perspectives (a)\,--\,(e).
Here, $60$ focus copies have been generated using the holographic approach from Ref.~\citenum{flamm2019beam} and distributed along a circular arc---please note the beam propagation direction parallel to the $z$-axis. Potential applications to this focus distribution could be surface texturing of curved articles, such as tubes, during a workpiece rotation.\cite{kumkar2017throughput} Here, parallel processing is enabled through the diffractive volume beam splitter and the fact that the MO is providing similar focus shapes and intensities for each of the split spots in the required working volume $V_{\text{W}} = \left( \unit[10]{\upmu m} \times \unit[200]{\upmu m} \times \unit[100]{\upmu m}\right)$. Uniformity errors amount to $<10\,\%$ (peak-to-valley). Depending on the available laser power performance and the processing strategy, the beam splitting can be further extended in the $x$-direction to reduce the number of passes or to increase the modification densities.\cite{kumkar2017throughput} 

\subsection{Volume-processing of transparent media}\label{sec:glassproc}
\begin{figure}[]
    \centering
    \includegraphics[width=1.0\textwidth]{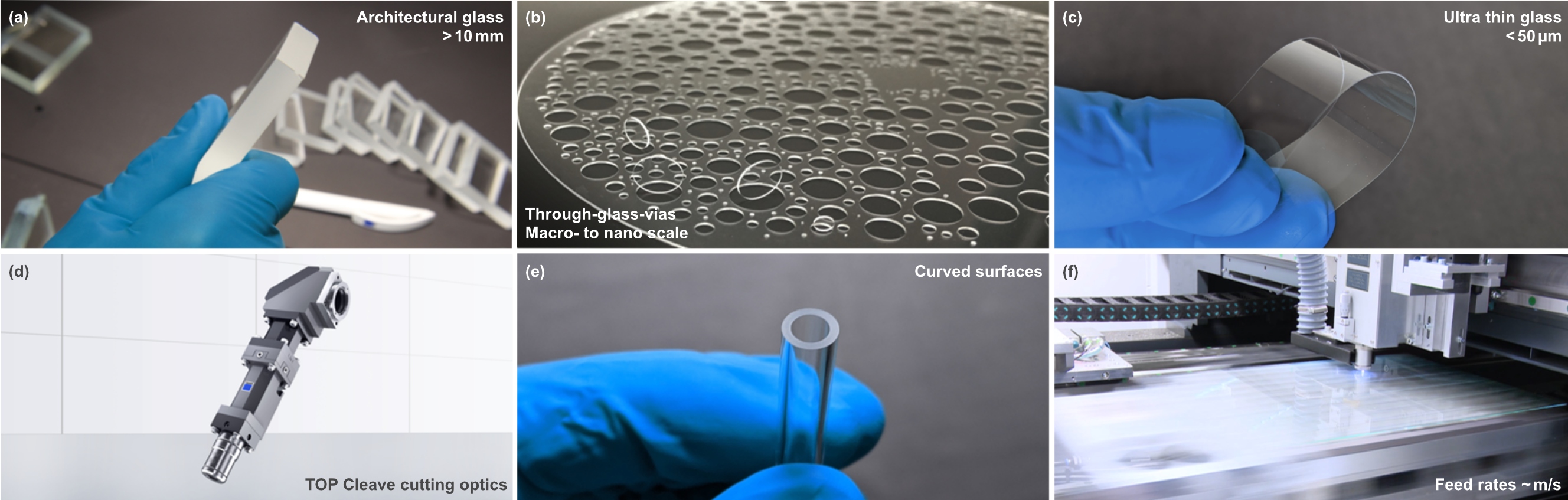}
    \caption{Selected glass cutting examples highlighting the performance of the \href{https://www.trumpf.com/en_SG/products/lasers/processing-optics/top-product-group/}{TOP Cleave cutting optics}. Cutting of $\unit[10]{mm}$-thick soda-lime glass using a high-energy version of a \href{https://www.trumpf.com/en_CA/products/lasers/short-and-ultrashort-pulse-laser/trumicro-series-5000/}{TruMicro Series 5000 laser} emitting $>\unit[1]{mJ}$ ultrashort laser pulses (a).\cite{jenne2020facilitated} Separating complex inner contours from TOP Cleave modifications and subsequent selective laser etching (b).\cite{kaiser2019selective} Controlled cutting of an ultrathin glass sample (thickness $<\unit[50]{\upmu m}$ using a TOP Cleave PRO processing head with crack control from preferential focus orientation and pulses from a \href{https://www.trumpf.com/en_INT/products/lasers/short-and-ultrashort-pulse-laser/trumicro-series-2000/}{TruMicro Series 2000 laser} (c).\cite{jenne2020facilitated} Photograph of a TOP Cleave cutting optics with visible modular concepts (bending mirror, beam expander, beam shaping unit and focusing unit (from top to bottom)) (d)\cite{flamm2021ultrafast}. Glas tube sample with diameter of $<\unit[8]{mm}$ processed with an adapted TOP Cleave cutting optics for curved interface compensation and subsequent thermal separation (e).\cite{rave2021glass} Example of glass panel processing with m/s feed rates (f). Please note the high-NA MO focusing the shaped pulses into the substrate and corresponding materials response (white plasma glow).\cite{bergner2018time, jenne2018high}}
    \label{fig:glass}
\end{figure}

Although not exclusively suitable for glass processing, the optical head shown in Fig.~\ref{fig:TC_gen} is very similar to the industrial glass cutting solution known as \href{https://www.trumpf.com/en_SG/products/lasers/processing-optics/top-product-group/}{TOP Cleave}. For this purpose the non-diffracting beam is length-adapted to the glass substrate by controlling raw-beam dimension and the magnification factor of the $4f$-like telescope, cf.~Sec.~\ref{sec:structuredmicro}. In combination with ultrashort pulses from TruMicro Series lasers, type-III-regime modifications\cite{itoh2006ultrafast} can be induced inside the bulk of the transparent article. During relative movement of the sample with respect to the optical head, a modified interface is generated acting as breaking layer.\cite{Flamm2015} A subsequent separation step strongly depends on the substrate (type of material, thickness), the target contour and the substrate shape. Typically, mechanical\cite{jenne2020facilitated}, thermal\cite{kaiser2023tailored} or chemical (selective laser etching)\cite{kaiser2019selective, hermans2014selective} separation strategies are applied to the ultrafast laser modified glass samples. A menu of glass articles with different thicknesses, shapes and contours is provided in Fig.~\ref{fig:glass}.
In any of the shown cases full-thickness modifications have been achieved from a single laser pass. The two extreme cases with substrate thicknesses of $>\unit[10]{mm}$ (a) and $<\unit[50]{\upmu m}$ (c) have been processed with an elliptical non-diffracting beam, similar to the one depicted in Fig.~\ref{fig:TC_gen}\,(c). Here, crack-control from asymmetric modifications facilitates the separation process.\cite{jenne2020facilitated, kleiner2023ultrafast} Potential feed rates amount to $\sim\unit[1]{m/s}$,\cite{flamm2021structured} see Fig.~\ref{fig:glass}\,(f), with typical surface roughnesses $S_a$ below $\unit[1]{\upmu m}$.\cite{rave2021glass} The optical head (d), is small, light and robust,\cite{flamm2021ultrafast} and compatible with fiber optical beam delivery based on hollow core fibers.\cite{baumbach2020hollow}
\par
\begin{figure}[]
    \centering
    \includegraphics[width=1.0\textwidth]{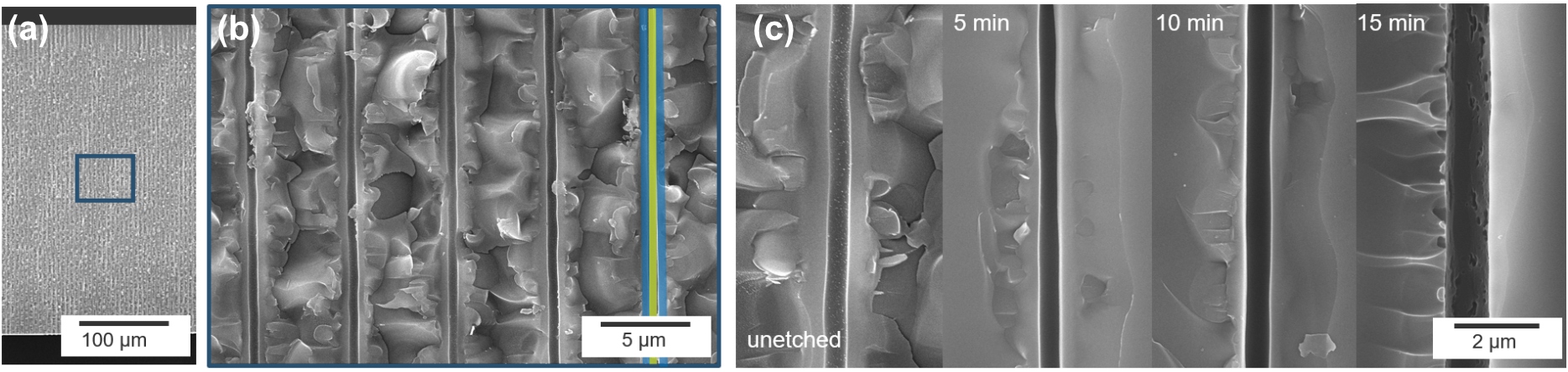}
    \caption{Microscope signals of a processed fused silica wafer exhibiting ultrafast laser induced nanochannels. Light microscope signal of the substrate's edge (a). Scanning electron micrograph of the nanochannels in different details (b), (c). Increasing the channel diameter by applying a selective laser etching strategy in a KOH solution (c).\cite{kaiser2019selective, flamm2021structured}}
    \label{fig:nann}
\end{figure}

 
We would like to emphasize that using our optical solutions for micromachining, cf.~Fig.~\ref{fig:TC_gen}, material modifications in the submicrometer regime, thus, nanomachining becomes possible. In Fig.~\ref{fig:nann}, nanochannels in a $\unit[350]{\upmu m}$-thick fused-silica wafer are shown exhibiting diameters of $\unit[500]{nm}$---an ideal starting for realizing interposers for high-density packaging.\cite{topper20103} The high aspect ratio hollow modifications show no taper angles and are achieved from applying single pulses.
A fine adjustment of the channel's diameter is enabled by an additional etching step. For example, as shown in Fig.~\ref{fig:nann}\,(c) a short etching time of a few minutes in a potassium hydroxide (KOH) solution allows the channels to be expanded to diameters of $>\unit[1]{\upmu m}$.\cite{flamm2021structured, kaiser2019selective}
\par
\begin{figure}[]
    \centering
    \includegraphics[width=.8\textwidth]{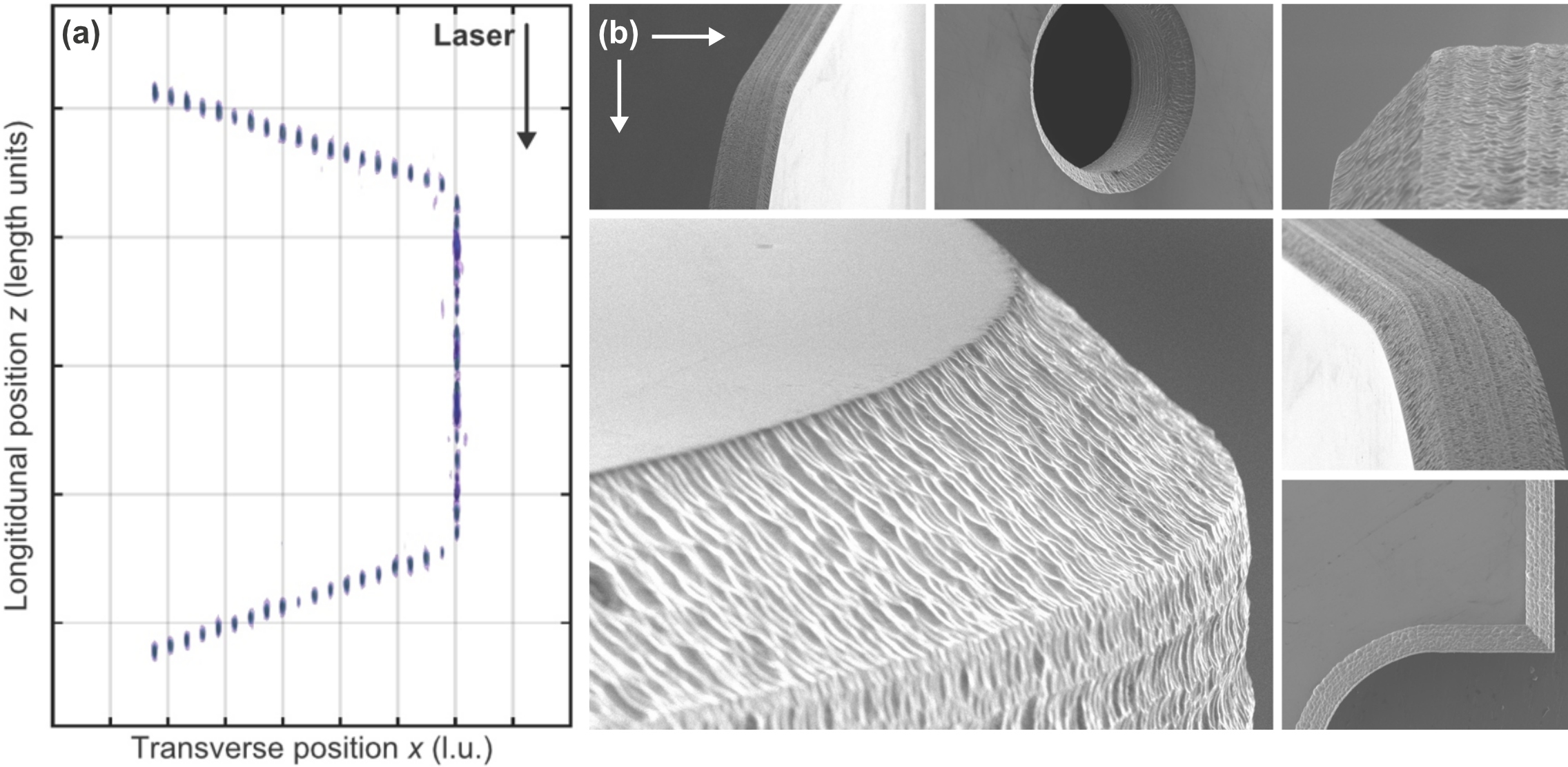}
    \caption{Tailored egde glass cleaving for edge-protecting applications.\cite{flamm2022protecting, marjanovic2019edge} A holographic 3D-beam splitter is distribution focus copies along an arbitrary edge shape, for example a 45$^\circ$ chamfer (a). After inducing volume-modifications from a \href{https://www.trumpf.com/en_INT/products/lasers/short-and-ultrashort-pulse-laser/trumicro-series-2000/}{TruMicro Series 2000 laser} (c).\cite{jenne2020facilitated} A selective etching strategy is applied to the glass substrate to achieve separation of complex inner and outer contours (b).\cite{kleiner2023laser}}
    \label{fig:champ}
\end{figure}

Recently, we have introduced photonic shaping tools\cite{flamm2021structured,flamm2022protecting, Flamm2022AOT} for scaling micromachining concepts in macroscopic working volumes of the focusing unit. As already discussed in Sec.~\ref{sec:structuredmicro}, the large working volume of the microscope objective is the main enabler of this approach. Prime application to this is the chamfered edge cutting of glass substrates, such as, e.g., display cover glasses.\cite{flamm2022protecting, kaiser2023tailored} Here, type-III-regime modifications\cite{itoh2006ultrafast} induced along arbitrary curved trajectories allows to generate chamfered or C-shaped edges from a single laser pass. The focus distribution takes the shape of the desired edge, see iso-intensity representation of the multi-spot focus in Fig.~\ref{fig:champ}\,(a), with a propagation direction parallel to the $z$-axis.
In this particular case, the holographic 3D-beam splitter is designed to process a $\unit[550]{\upmu m}$-thick Corning\textsuperscript{\textregistered} Gorilla\textsuperscript{\textregistered} glass sample with $\unit[100]{\upmu m}$, $45^\circ$ bevels on both interfaces. Now, edge protection is enabled as the substrate's edge angle is reduced from a straight edge to $\approx 45^\circ$, see Fig.~\ref{fig:champ}\,(b).\cite{marjanovic2019edge, bukieda2020study, flamm2022protecting} The separation along complex inner and outer contours, see vias and notches in (b), is achieved from applying a selective laser etching strategy.\cite{gottmann2017selective, kaiser2019selective, kleiner2023laser} Please note, that the optical tool enables processing in a single pass---a workpiece flipping is not necessary.\cite{flamm2022protecting}

\section{Modifications in the millimeter scale}\label{sec:milli}
Laser material modifications in the millimeter scale, include a variety of laser processes and applications, such as laser cutting, welding, marking, and cleaning. \textcolor{black}{Relative movement between workpiece and laser radiation can be achieved by axis stages---however, the most versatile tool for such applications represents a laser scanning optics with remarkable features such as high accuracy, dynamic functionality and modularity}\cite{lu2013review,yurevich2015optical}. In this section, we will discuss the unique characteristics and challenges for scanning optics in the realm of laser remote welding applications. This field typically covers focal spot diameters from a few micrometer up to the lower millimeter range, along with scanning field diameters extending from a few millimeters up to the meter range. An integral component of such an optical setup is the focusing optics---an F-theta lens in our configuration. We start by explaining its design in Sec.~\ref{sec:FTheta}, moving on to \textcolor{black}{emphasize} its importance in combination with beam shaping Sec.~\ref{sec:FTBS}. Finally, we will demonstrate a high power beam shaping application in Sec.~\ref{sec:FTApp} to \textcolor{black}{demonstrate} the beneficial interaction of all components.

\subsection{Large scanning field F-Theta lens}\label{sec:FTheta}
Material modifications at the millimeter scale involve common design considerations similar to focusing units for micromachining, cf.~Sec.~\ref{sec:MO}, including low aberration designs and high power suitability. However, the significance of field-dependent aberrations is typically increased due to large scanning fields compared to the focal length $f$. This necessitates achieving planar imaging fields and ensuring consistent imaging performance throughout the scanning field.\cite{ozga2018unseen, yurevich2015optical} 
\par
Considering the widely used pre-objective scanning configuration, where the scanning unit is positioned in front of the focusing unit,\cite{marshall2012handbook} the objective typically consists of multiple lenses for a flat-field design\cite{marshall2012handbook}. 
Specifically, F-Theta lens designs are meticulously optimized to generate uniform focal shapes throughout the scanning field while minimizing field curvature.  
\par
In the context of laser remote welding, scanning fields commonly range  from approximately $\unit[100]{mm}$ to $\unit[1000]{mm}$. \textcolor{black}{Given that achieving homogeneous and repeatable processing results is mandatory, any variations along the propagation direction can be set in relation to the Rayleigh length $z_{\text{R}}$.} Consequently, effects like field curvature, astigmatism, or thermal focal shift \textcolor{black}{have to be smaller than one Rayleigh length.} Telecentricity and field-dependent aberrations, including astigmatism and coma, must also be considered. This necessitates a balance in the optical design strategy, compromising between a low aberration design, which demands a higher number of lenses, and a thermally stable, cost-effective design with fewer lenses. Integrating these design considerations with additional complexities arising from advanced beam shaping elements mandates a comprehensive system simulation, as discussed in the subsequent section~\ref{sec:FTBS}.
\par
\begin{figure}[]
    \centering
    \includegraphics[width=0.9\textwidth]{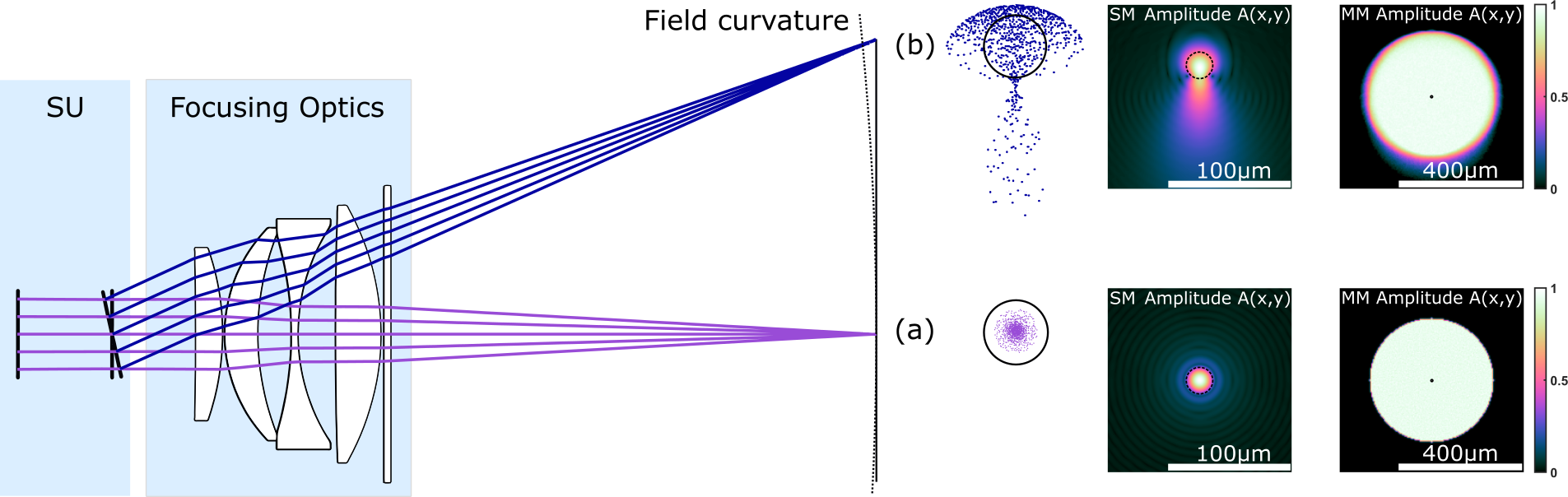}
    \caption{Ray optical layout of a F-Theta lens with integrated scanning unit and common aberrations. The purple rays illustrate an undeflected field, showcasing diffraction-limited imaging performance, as indicated by the airy radius (black circle) enveloping all traced rays. Subsequently, a wave optical simulation illustrates the aberration impact on single-mode source, while a convolution approach is applied to a multi-mode source~(a). In contrast, the blue rays represent a field at maximum deflection angle, showing the influence of aberrations through a non-diffraction-limited spot diagram and field curvature. Subsequently, a wave optical simulation illustrates the aberration impact on a single-mode source, while a convolution approach is applied to a multi-mode source~(b).     }
    \label{fig:FTLayout}
\end{figure}

Although the field of application for F-Theta lenses is large, here, we focus on high power laser remote welding in the $\unit[\left(1\dots20\right)]{kW}$ domain. For the F-Theta lens, the most common requirements on focal length is within $\unit[200]{mm}$ and $\unit[500]{mm}$ with scan field diameters of $\left(0.5\dots1.0\right)$ times the focal length and numerical apertures of $\left(0.02\dots0.1\right)$. 
As emphasized earlier, highly brilliant laser sources, particularly single-mode fiber lasers such as \href{https://www.trumpf.com/en_US/products/lasers/fiber-laser/trufiber-p/}{TruFiber P series}, necessitate superior imaging performance to minimize aberrations and ensure negligible impact on the beam profile. 
During the design phase of an F-Theta lens, the evaluation typically \textcolor{black}{focuses on}  diffraction-limited imaging performance. This approach implies that the lens system's performance is predominantly constrained by diffraction effects rather than other optical aberrations or imperfections. Quantitative criteria for this evaluation include the Strehl ratio, commonly set at $\geq 0.8$, corresponding to a $\lambda$/4 peak-to-valley wavefront error (Rayleigh’s criterion) for low-order aberrations, and a $\lambda$ /14 root mean square wavefront error (Maréchal criterion). \cite{sun2016lens, williams2002introduction, marechal1948etude,gross2005handbook}
In the early design phase, when dealing with larger aberrations, evaluating spot diagrams provides a reliable estimate of transverse geometrical aberration and its impact compared to a diffraction-limited design.\cite{gross2005handbook} This comparative evaluation is illustrated in the exemplary F-Theta lens design representation shown in Fig.~\ref{fig:FTLayout}\, which highlights common aberration effects in F-Theta lens designs. 
\par

The design features a scanning unit (SU) followed by the focusing optics based on a four-lens design \cite{hellstern2022patent}. \textcolor{black}{The lens group}, with an effective focal length of $f=\unit[270]{mm}$, a maximum field diameter of $d_{\text{F}}=\unit[260]{mm}$, and a focusing numerical aperture NA$=0.055$ at a wavelength range of $\unit[\left(1030\dots1100\right)]{nm}$, is an optimized F-Theta lens design with large scanning field diameter ($d_{\text{F}}/f\approx0.96$) for high power applications. F-Theta lenses are predominantly derived from photographic lens designs, \textcolor{black}{using one half} of a Double Gauss lens.\cite{yurevich2015optical} The four-lens configuration depicted in Fig.~\ref{fig:FTLayout} deviates from a Cook triplet design, characterized by two positive lens groups surrounding a negative lens group. The advantage of this design lies in a large scanning field while maintaining a compact lens design and high imaging performance---\textcolor{black}{specifically, a flat field and low aberrations at the outer edge of the scanning field.}
\par
As previously mentioned, the influence of aberrations on the focal spot can significantly affect processing outcomes. Consequently, we will examine the field-dependent aberrations illustrated in Fig.~\ref{fig:FTLayout} and assess their impact on the beam profile for both a single-mode source and a multi-mode source. This analysis underscores the critical importance of tailoring lens design to application-specific requirements. 
The purple rays in the illustration depict the focusing of an undeflected field at the center of the scanning field. Despite the anticipation of dominant spherical aberrations, the spot diagram reveals that all rays fall within the $\unit[22.8]{\upmu m}$ Airy disk, indicating diffraction-limited performance. A more suitable criterion to evaluate the impact is the Strehl ratio, and for an even more detailed evaluation, a wave optical simulation of the system. Such wave optical simulation of a single-mode illumination (SM) is presented in the subsequent \textcolor{black}{amplitude distribution} in the focal plane. This simulation indicates minimal impact ($M^{\text{2}}\approx1.1$, $I_{\text{peak}}=1$). Multi-mode sources (MM), such as \href{https://www.trumpf.com/en_US/products/lasers/disk-lasers/trudisk/}{TruDisk} or \href{https://www.trumpf.com/en_US/products/lasers/fiber-laser/trufiber-s/}{TruFiber S}, featuring a $\unit[100]{\upmu m}$ fiber diameter and emitting a flat-top-like intensity distribution ($SPP=\unit[4]{mm \times mrad}$), can be considered as partially incoherent. The emitted  A straightforward approach involves convolution of the fiber output and the point spread function.\cite{Wolf2014-fw} This convolution result is illustrated in the subsequent multi-mode image, \textcolor{black}{demonstrating the impact of aberrations on beam} characteristics ($I_{\text{peak}}=1$).
\par
However, at the maximum deflection angle, the impact of multiple field dependent aberrations is demonstrated in the blue field shown in Fig.~\ref{fig:FTLayout} . Notably, the presence of field curvature is evident, as indicated by the focal position in front of the working plane. It is crucial for this difference to be significantly smaller than the Rayleigh length to ensure uniform processing results throughout the scanning field, particularly in sensitive applications such as gas-tight welding.\cite{moller2022spatially} As mentioned before, the spot diagram already gives some indication of present aberrations. The shown distribution is indicative for a combination of astigmatism and coma, both being field-dependent aberrations as expected. 
The potential influence on the beam profile can be estimated by comparing the rms-radius of the spot diagram to the Airy radius, resulting in a ratio of 1.64.
Consequently, the system is not diffraction-limited, suggesting an anticipated change in the beam profile for single-mode illumination. This expectation is confirmed by the single-mode source simulation, revealing significant differences from the undeflected field ($I_{\text{peak}}\approx0.4$), \textcolor{black}{most likely causing} process inconsistencies. Conversely, the effect on extended multi-mode sources is observed to be minimal( $I_{\text{peak}}\approx0.98$)---most likely irrelevant for many applications.
\par
The design analysis highlights the absence of a universal approach for focusing optics. Instead, design considerations should align with the specific field of application and laser sources. While diffraction-limited lens designs are crucial for precision in applications with highly brilliant laser sources, they entail increased costs and potential high-power limitations. On the other hand, non-diffraction-limited designs offer cost-effective alternatives with greater flexibility, although they may encounter constraints with brilliant sources and \textcolor{black}{in critical} applications. A comprehensive evaluation, considering factors such as intended applications, aberration budget, compactness, laser sources, power, and costs, remains crucial for tailoring the design to its specific context.

\subsection{Beam shaping concepts for Scanning optics}\label{sec:FTBS}
Improving the processing speed and, consequently, enhancing system productivity constitutes a primary objective in the domain of laser remote welding. Sustaining this progress mandates the integration of highly brilliant, high-power laser sources, coupled with advanced beam shaping solutions \cite{moller2022novel,moller2022spatially}. Tailoring the laser beam's shape to specific process requirements enables precise control of heat input, mitigating adverse effects like thermal distortion, stress, and undesirable metallurgical transformations. This customization enhances penetration depth, minimizes spatter, and boosts overall process efficiency \cite{moller2022novel,moller2022spatially}. Fiber-based beam shaping technologies, such as \mbox{2-in-1} fibers (TruDisk platform: \href{https://www.trumpf.com/filestorage/TRUMPF_Master/Products/Lasers/02_Brochures/TRUMPF-disk-laser-BrightLineWeld-Flyer-DE.pdf}{BrightLine Weld} and TruFiber platform: \href{https://www.trumpf.com/de_DE/newsroom/pressemitteilungen-global/pressemitteilung-detailseite-global/release/messe-laser-trumpf-praesentiert-besonders-vielseitigen-faserlaser-7919/}{BrightLine Mode}) or beam shaping elements within processing optics, facilitate the creation of tailored intensity distributions (e.g., multiple spots\cite{kumkar2017throughput,moller2022novel,moller2022spatially} or ring shapes\cite{grunewald2021influence}). It is noteworthy that these beam shaping solutions impose additional criteria on the F-Theta lens design such as enhancing the focusing \textcolor{black}{NA} and demanding minimal distortion. Moreover, the precise arrangement of the fiber, beam shaping elements, and focusing optics plays a critical role in manipulating and optimizing such focal distributions. Achieving optimal performance involves considerations of the interplay among fibers, beam shaping elements, and focusing optics on a holistic optical design, thereby enhancing, or constraining the propagation behavior of focus distributions. \textcolor{black}{Emphasizing the importance of a holistic optical design approach for achieving optimal laser remote welding outcomes.}
\par
\begin{figure}[]
    \centering
    \includegraphics[width=0.95\textwidth]{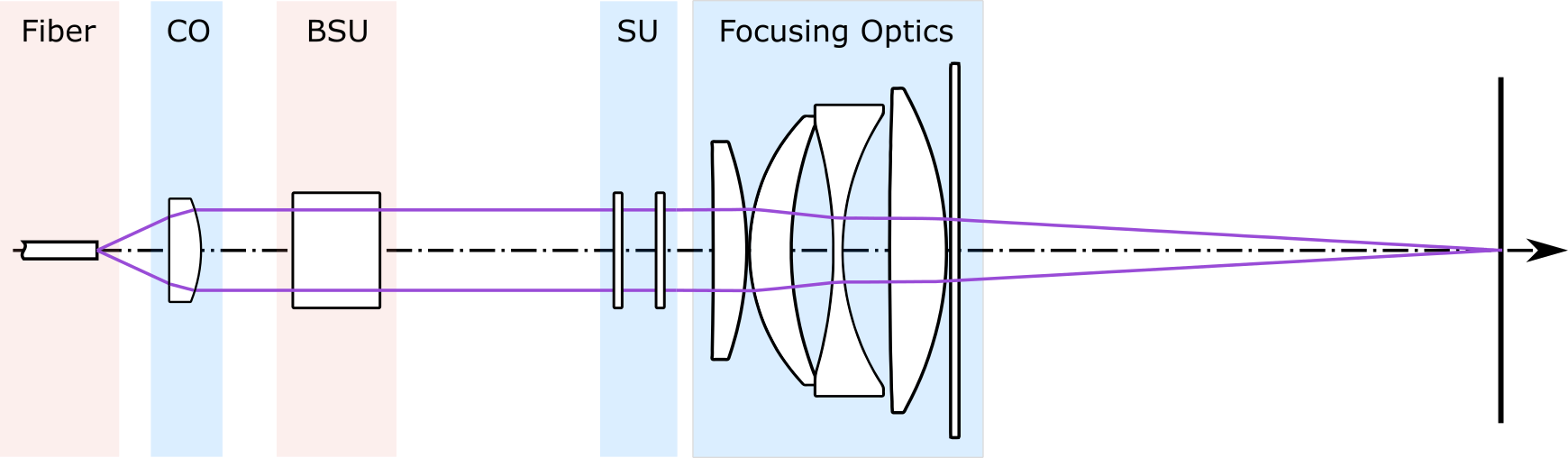}
    \caption{Optical arrangement of scanning optics designed for beam shaping applications. Modules on red background offer great potential for beam shaping. The fiber output is collimated by the collimation optics (CO). The collimated beam then propagates to the beam shaping unit (BSU), allowing for alterations to the beam's amplitude, phase, and/or polarization. Subsequently, a scanning unit (SU) deflects the beam towards the focusing optics. The interaction with the focusing optics induces a spatial movement of the focused spot, allowing a dynamic control of the laser beam.}
    \label{fig:SOLayout}
\end{figure}
Scanning optics present various options for integrating beam shaping into the setup. Here, we explore two prominent implementation strategies. The schematic of a scanning system is depicted in Fig.~\ref{fig:BSLayout}, with the fundamental \textcolor{black}{modules} starting from the divergent fiber output. This output undergoes collimation optics, and the resulting collimated beam is deflected within the scanning unit, leading to a spatial displacement on the working plane, cf. Sec.~\ref{sec:FTheta}.
The beam profile in the working plane can be estimated as an aberrated, magnified image of the source output field. This description can be analytically simplified to a $4f$-arrangement, expressed as \textcolor{black}{$\mathcal{F}\{\mathcal{F}\{f_{\text{fiber}} \left(x,y\right)\}\}= f_{\text{wp}}\left(-x,-y\right)$}, where $\mathcal{F}$ represents the Fourier transform. The complex field in the working plane $f_{\text{wp}}\left(-x,-y\right)$, acts as the inverse replica of the input field at the fiber $f_{\text{fiber}}\left(x,y\right)$.\cite{goodman2005introduction} Typically, in a good approximation beam profiles are considered as rotational symmetrical Gaussian-like shapes for single-mode sources and top-hat-like shapes for larger multi-mode fibers.
This imaging relation underscores the potential and impact of the fiber output field on the focal distribution. Consequently, an evident beam shaping approach arises from the fiber geometry. In addition to the mentioned round fibers, polygonal fiber geometries, fiber arrays~\cite{zhu2007birefringent}, and notably, multi-core fibers, such as a 2-in-1 fiber, offer a direct impact on beam shaping. Dynamic beam shaping is also achievable, for instance, by altering the \textcolor{black}{power distribution between the inner core and outer core of a 2-in-1 fiber.}
The second implementation can occur within the collimated beam, as demonstrated by the beam shaping unit. The focusing optics performs a far-field transformation, ideally expressed through a $2f$-arrangement as $\mathcal{F}\{f_{\text{BSU}} \left(x,y\right)\}\propto f_{\text{wp}}\left(x,y\right)$, cf. Sec.~\ref{sec:structuredmicro}. This setup allows for the implementation of various polarization, amplitude, and notably, phase manipulations. Common techniques involve phase manipulations through refractive or diffractive beam shaping elements, such as beam splitters\cite{kumkar2017throughput}, flat-top shapers, and ring shapers, to modify the output field at the working plane. The potential for beam shaping expands significantly when combining the described strategies at both the fiber and beam shaping unit.

\subsection{Material processing in e-mobility applications}\label{sec:FTApp}
The field of applications for laser remote welding is extensive, with emerging laser technologies continually exploring novel avenues. A significant frontier in this landscape is the realm of e-mobility, characterized by challenging applications and a high demand for tailored beam shaping to enhance productivity and process stability.\cite{ziegler2022business} To underscore the significance of beam shaping in augmenting productivity, we explore the optical design challenges of a specific application, namely the gas-tight welding of aluminum alloys for cooling plates and casted housings.\cite{moller2022novel, moller2022spatially} A primary objective in this context is the reduction of pores and cracks at elevated welding speeds, necessitating the establishment of a stable keyhole.
\par
\begin{figure}[]
    \centering
    \includegraphics[width=1.0\textwidth]{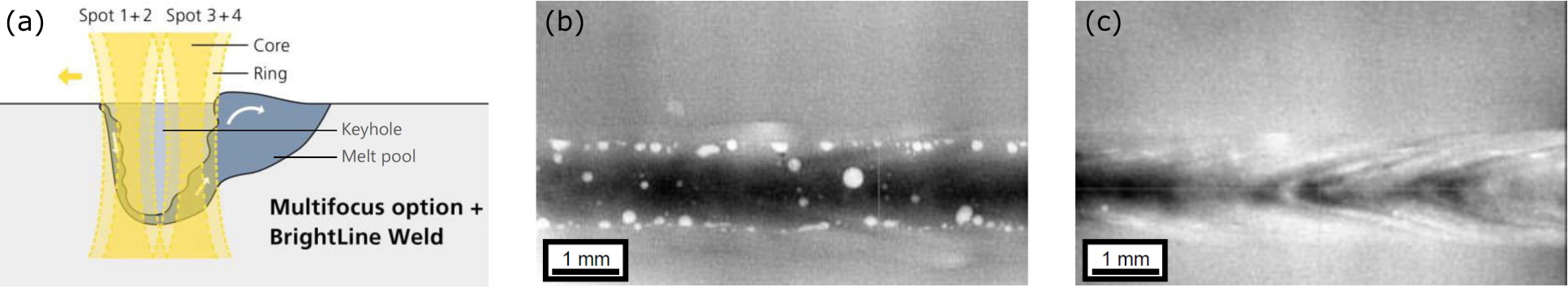}
    \caption{Gas-tight welding of aluminum alloys performed on a \href{https://www.trumpf.com/en_US/products/machines-systems/laser-welding-systems-and-the-arc-welding-cell/trulaser-cell-3000/}{TruLaser Cell 3000} equipped with a \href{https://www.trumpf.com/en_US/products/lasers/disk-lasers/trudisk/}{TruDisk 8001} laser source operating at 8kW.\cite{moller2022novel, moller2022spatially} Stabilization of the keyhole is achieved by distributing four focal points arranged in a tailored quadratic configuration, resulting in the enlargement of the keyhole and reduction in its fluctuation~(a). Pores, evident as white dots in the X-ray image resulting from single-beam welding using the 2-in-1 fiber technology \href{https://www.trumpf.com/filestorage/TRUMPF_Master/Products/Lasers/02_Brochures/TRUMPF-disk-laser-BrightLineWeld-Flyer-DE.pdf}{BrightLine Weld}~(b). The implementation of multi-foci beam shaping approach results in a notable reduction in porosity~(c). \cite{moller2022novel, moller2022spatially}
    }
    \label{fig:Keyhole}
\end{figure}
\begin{figure}[]
    \centering
    \includegraphics[width=0.95\textwidth]{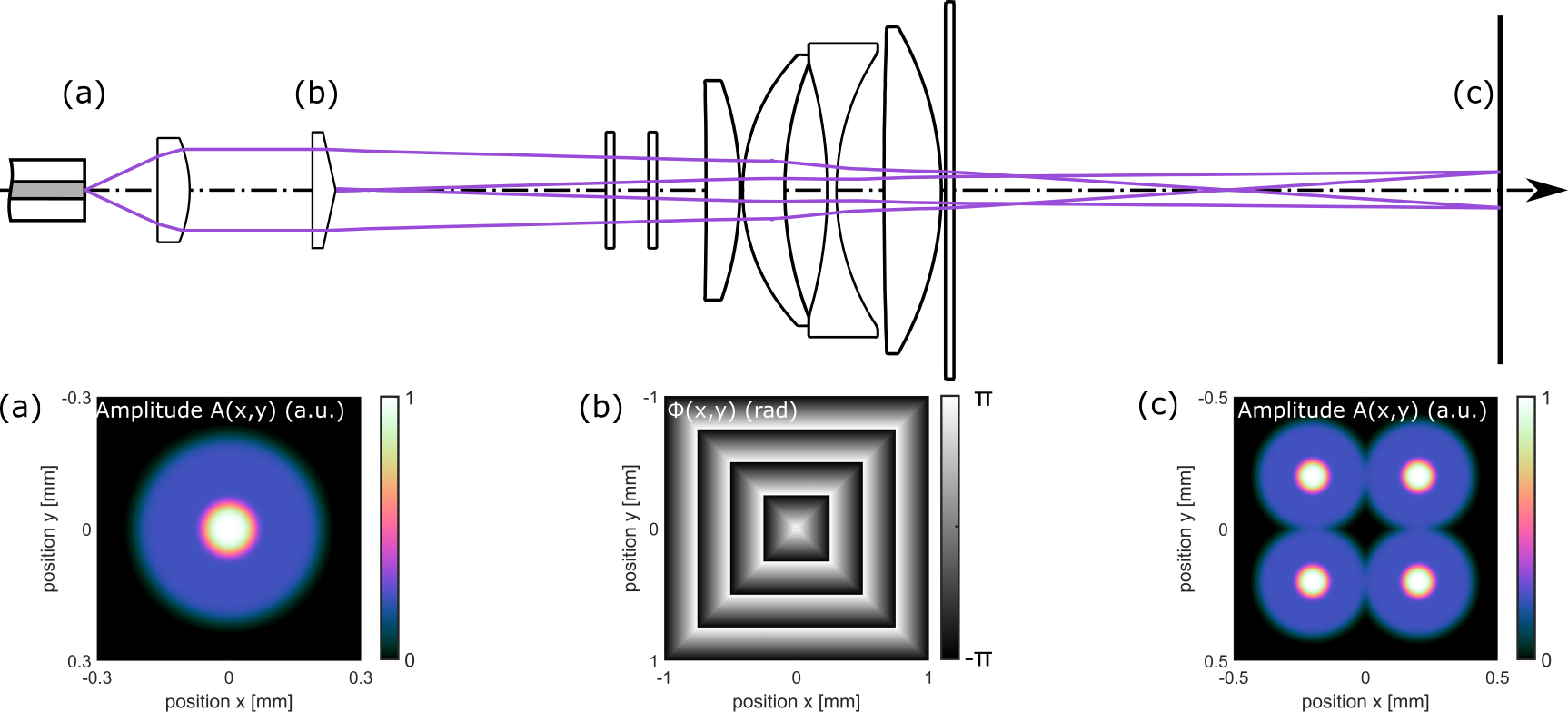}
    \caption{Exemplary multi-spot beam shaping configuration for gas-tight welding. The power ratio of the 2-in-1 fiber ($\unit[50]{\upmu m}$ inner core diameter and $\unit[200]{\upmu m}$ outer core diameter) is adjustable to attain an output field resembling a).  Additionally, the introduction of beam shaping optics in the collimated beam ($f = \unit[140]{mm}$), such as a pyramidal phase element b), achieves a beam-splitting function. Post the focusing unit ($f = \unit[270]{mm}$, $\text{NA} = 0.05$), both beam shaping implementations collaboratively operate to generate four magnified ($m=1.9$) images of the fiber output, arranged in a square pattern ($\unit[390]{\upmu m} \times \unit[390]{\upmu m}$ ) within the focal plane c). }
    \label{fig:BSLayout}
\end{figure}

In Figure~\ref{fig:Keyhole}\, we present a schematic depicting such a stable keyhole, achieved through the implementation of a beam shaping concept employing multiple foci distribution. Möller \textit{et al.}~(Ref.~\citenum{moller2022novel}) demonstrated a remarkable enlargement and stabilization of the keyhole size, resulting in 7\,\% size variance compared to the substantial 53\,\% observed in single-beam processing. The proposed focal distribution comprises four equally weighted foci arranged in a quadratic configuration. The realization of this focal distribution involves the use of a 2-in-1 fiber to define the single spot distribution, while a beam splitting element within the beam shaping unit generates four identical foci arranged in a square pattern, as illustrated in Fig.~\ref{fig:BSLayout} and detailed in Sec.~\ref{sec:FTBS}.

The 2-in-1 fiber output distribution is dynamically adjustable by altering the power ratio between the inner and outer core.\cite{haug2019beam} This adjustment results in the illustrated amplitude distribution shown in Fig.~\ref{fig:BSLayout}~(a). The distribution can be specifically tailored to the process or material, enabling the modification of the heat-affected zone, thereby allowing for higher feed rates, reduced spatter, and enhanced process stability.\cite{haug2019beam}
The fiber output field is then imaged onto the working plane using a $4f$-like setup consisting of collimation and focusing optics. The far-field distribution of the fiber at the beam shaping unit is modified by a beam-splitting element, such as inducing a pyramidal phase function Fig.~\ref{fig:BSLayout}~(b). The resulting focal distribution in the working plane forms an array-like pattern of four identical images of the fiber output field, see Fig.~\ref{fig:BSLayout}~(c). The distance between the focal spots is determined by the induced deflection angle of the beam-splitting element and the effective focal length of the focusing optics. Adjustment of either of these parameters allows for the customization of the spot distance.
As demonstrated in Section~\ref{sec:FTheta}, the field-dependent aberrations of the scanning optics must be considered. Therefore, a comprehensive system design is strongly recommended to avoid process inconsistencies and achieve a cost-effective system.
\par
In recent investigations, we demonstrated the significant importance and broad range of applications for scanning optics. Consequently, the thoughtful consideration of design aspects in focusing optics emerges as a pivotal factor in achieving consistent processing outcomes (cf. Sec.~\ref{sec:FTheta}). Figure~\ref{fig:FTLayout} illustrates a critical aspect of the design process, showcasing the interplay between managing manufacturing costs and \textcolor{black}{reasonable} budgeting for aberrations to meet process-dependent requirements.
To further enhance productivity and explore novel applications, beam shaping plays a pivotal role, as elaborated in Sec.~\ref{sec:FTBS}. The customization of focal distribution to align with specific processes and materials is crucial. An illustrative example beside the discussed gas-tight welding of aluminium alloys\cite{moller2022novel} is the enhancement of sheet steel engraving productivity through multi-spot parallel processing using ultra-short pulse laser (\href{https://www.trumpf.com/en_US/products/lasers/short-and-ultrashort-pulse-laser/trumicro-series-5000/}{TruMicro 5070}).\cite{flamm2019beam, kumkar2017throughput} The discussion in Section~\ref{sec:FTBS} elaborates on combined beam shaping strategies involving fiber and a beam shaping unit, revealing significant potential in tailoring the focal distribution to meet precise process requirements.

\section{Modifications in the meter scale}\label{sec:meter}

When large area processing with uniform intensity distribution is required, line beam systems represent the laser tool-of-choice enabling processing with both highest uniformities and intensities. Applications include the post-treatment of architectural glass and process steps in display manufacturing such as laser lift-off (LLO) \cite{LLO} or solid state laser annealing (SLA) \cite{SLA, 10.1117/12.2608238 }. We provide a flexible platform based on diode pumped solid state laser technology (DPSSL) coupled with advanced beam shaping techniques. Depending on processing parameters and productivity requirements of specific applications, our optical design can be modified to different line lengths and short axis parameters. In the following, we discuss our imaging system (Sec.~\ref{sec:img_m_scale}) and present beam shaping techniques (Sec.~\ref{sec:LBS}) which are necessary to assure appropriate line-beam specifications. Moreover, the range of application can be extended considerably by modulating the line beam. An interesting application can also be identified here in the display industry, namely laser-induced forward transfer (LIFT)\cite{serra2019laser}, in which a structured line beam offers a scalable approach for mass transfer (see Sec.~\ref{sec:multispot}). 

\subsection{Focusing/imaging on meter scale}\label{sec:img_m_scale}

\begin{figure}[]
    \centering
    \includegraphics[width=0.9\textwidth]{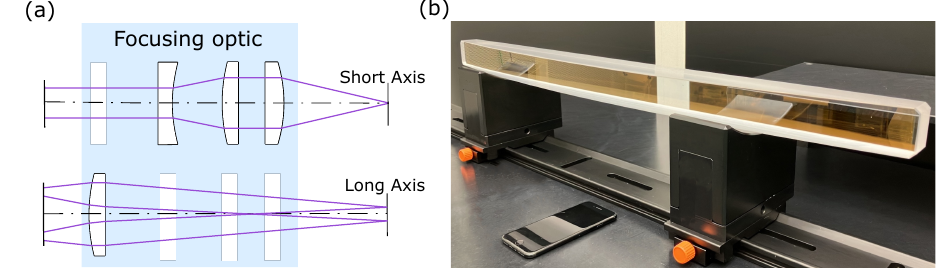}
    \caption{Exemplary layout of a focusing unit of a line beam system (a). The optical layout can be described in two independent axes, a short axis focusing unit (top), and a long axis Fourier system (bottom). Picture of a large optic from a UV line beam system to demonstrate dimension. The lens is made from a monolithic block of fused silica (b).}
    \label{fig:largeOptic}
\end{figure}

A wide field of line beam dimensions can be covered based on our platform. For example, UV line beam systems with lengths up to $\unit[1500]{mm}$ have been realized, whereas up to $\unit[3200]{mm}$ have been demonstrated with the modular stitching concept of the \href{https://www.trumpf.com/de_DE/produkte/laser/kurz-und-ultrakurzpulslaser/linienoptik-infrarot/}{IR line}. The latter is based on the \href{https://www.trumpf.com/en_US/products/lasers/disk-lasers/trudisk/}{TruDisk series}, where several fiber-coupled line beam modules are combined to processing with more than $\unit[100]{kW}$ in average power. It is used for rapid thermal annealing of thin films \textcolor{black}{deposited} on large architectural glass, cf.~Sec.~\ref{sec:multispot}\cite{IRLinie}. Due to the demands on highest uniformity as well as strict requirements for angular spectrum of the line beam, the UV line beam system, in contrast to the IR version, consists of a monolithic imaging system. Consequently, the imaging system comprises optical lenses within the meter range, see Fig.~\ref{fig:largeOptic}\,(b). This brings inherent challenges in the entire production chain of these large optics due to its immense size and considerable weight. Starting with optical specifications, through manufacturing, characterization, handling and finally a proper mounting concept in the line beam system. For instance, distinct design considerations which are beyond standard optical design norms such as ISO 10110-6 \cite{ISO10110-6} are required. Material properties such as the homogeneity of the refractive index, glass- or surface imperfections and homogeneity of coatings are challenging requirements for manufacturing within this size regime. In large optics manufacturing, a tool set for measurement and analysis has been established to guarantee absolutely reliable optical quality at the limits of technical feasibility. A sophisticated process chain ensures that the quality of these sensitive components can be maintained through the entire product life cycle. Highest requirements in surface properties and in combination with non-neglectable weight of these large components is taken into account by the mechanical design of the lens mount that minimizes the negative effects of gravity.
\par Due to the strong asymmetrical aspect ratio of the line focus, the imaging design is based on cylindrical lenses. Typically optical properties are described in two independent axes, which are denoted by long and short axis. Figure~\ref{fig:largeOptic} (a) shows an exemplary layout of such a focusing unit, where the short axis is illustrated on the top and the long axis on the bottom. The imaging system in short axis is based on a lens-triplet design and is aberration-optimized to guarantee diffraction limited imaging quality along the line beam.  
\par 
 
\subsection{Shaping the line focus}\label{sec:LBS}
\begin{figure}[]
    \centering
    \includegraphics[width=1\textwidth]{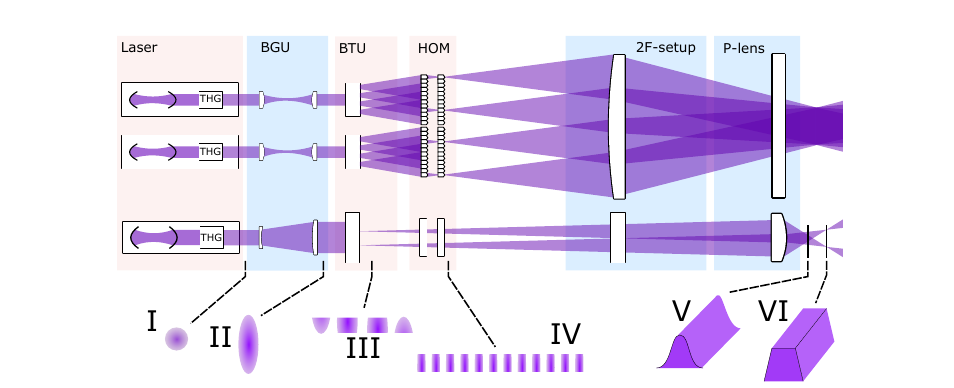}
    \caption{Optical layout of a UV line beam system. Long axis beam shaping (top), short axis beam shaping (bottom). The laser source consists of an infrared resonator and a third harmonic generator (THG). The latter emits a roundly shaped beam (I) which is shaped into a elliptical beam (II) in a subsequent beam guiding unit (BGU). The beam transformation unit (BTU) cuts the beam into several beam-lets (III), which illuminate an imaging homogenizer (HOM). A $2f$-setup superimposes the output packages (IV) in the working plane. In short axis, a projection system either focuses or images the BTU output to the working plane, resulting in a Gaussian (V) or flat-top (VI) intensity distribution \cite{10.1117/12.2608238}.}
    \label{fig:LFS}
\end{figure}

In this section, the beam shaping concept of a UV line beam system is presented. Figure~\ref{fig:LFS} depicts the individual segments of the optical concept briefly, however does not cover technical details of our optical design. 
\par
The beam source of a UV line beam system is based on the \href{https://www.trumpf.com/en_US/products/lasers/short-and-ultrashort-pulse-laser/trumicro-series-8000/}{TruMicro Series 8000} platform. This laser technology delivers superior reliability and low maintenance costs combined with very high pulse energy stability. It consists of a diode pumped, infrared resonator and a third harmonic generation (THG). The latter delivers partially coherent, roundly shaped laser beams at wavelength of $\unit[343]{nm}$ with beam quality of $M^2 \approx 25$. To ensure appropriate line beam specifications, special beam shaping methods are essential which will be presented in the following. Starting with the beam guiding unit (BGU), which utilizes fundamental anamorphotic beam shaping techniques to shape and guide elliptical beams (II) to the beam transformation unit (BTU). The BTU manipulates the beam quality in the long and short axes, resulting in degraded beam quality along the long axis. This reduces interference effects, which might be caused by homogenizing techniques. In the short axis, the beam quality is improved which enhances focusability \cite{10.1117/12.2608238}. The beam profile (III) in Fig.~\ref{fig:LFS}
shows a typical output of the beam transformation unit. An initially elliptical beam (II) with similar beam qualities in both axes is separated into a number of beam-lets which are spatially reoriented such that a beam quality ratio of $M^2_{x}>>M^2_{y}$ is realized \cite{10.1117/12.2608238}. After the BTU, the system can be split into two parts: a long axis (LA) and a short axis (SA) beam shaping (see Fig. \ref{fig:LFS} top/bottom). In long axis the beam-lets overlap and smoothly illuminate an imaging homogenizer (HOM). The latter splits the beam into several beam-lets in long axis (IV), which are projected by a Fourier system configuration ($2f$-setup), in a far field plane. The result is a homogeneous flat-top shaped beam profile in long axis\cite{10.1117/12.2608238}. In short axis, the beams pass through a lens system which either image or focus the BTU output into the working plane. This results in either a flat-top or a Gaussian shaped short axis profile (see beam profiles (V) and (VI) in Fig.~\ref{fig:LFS}). 
\par
\begin{figure}[]
    \centering
    \includegraphics[width=0.95\textwidth]{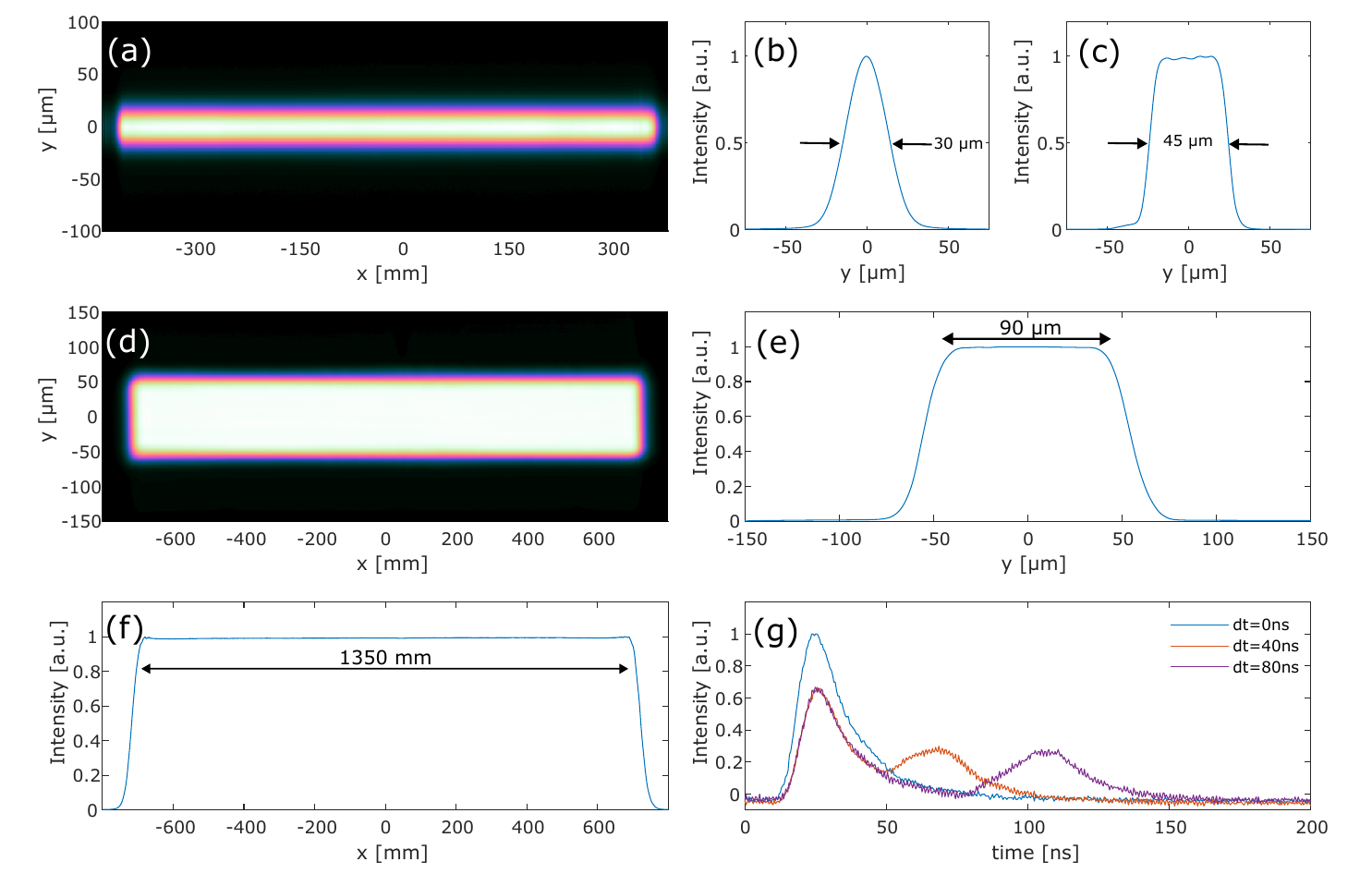}
    \caption{Measurements of UV line beam systems at maximum fluence level with different line lengths ($\unit[750]{mm}$ and $\unit[1350]{mm}$ (a, d)) and the corresponding short axis intensity distributions. Within the $\unit[750]{mm}$ system, the short axis can be either formed to a Gaussian profile with $\unit[30]{\upmu m}$ width (b) or a flat-top profile with $\unit[45]{\upmu m}$ (c). The short axis profile (e) of the $\unit[1350]{mm}$ has a width of $\unit[90]{\upmu m}$ ($FW$ at $90\,\%$) and a standard deviation of the intensity distribution of less than 2.5$\,\%$, whereas the long axis profile (f) has a homogeneity of less than $1.5\,\%$ (both $2\sigma$ at $96\,\%$). This measurement was performed with eight synchronized beam sources (\href{https://www.trumpf.com/en_US/products/lasers/short-and-ultrashort-pulse-laser/trumicro-series-8000/}{TruMicro Series 8000}). Examples of temporal pulse shaping by synchronizing multiple lasers and shifting by specific delays (i.e. $dt = \unit[40]{ns}$ or $dt = \unit[80]{ns}$) to achieve a two hump profile (g).}
    \label{fig:lineBeamMeasurement}
\end{figure}

 
Short axis width is a recurrent trade off between necessary heat input for a specific process and economical aspects such as processing speeds. The platform is prepared to deliver great flexibility, widths vary between $\unit[30]{\upmu m}$ and $\unit[100]{\upmu m}$. 
In Fig.~\ref{fig:lineBeamMeasurement} (a--g), exemplary measurements exhibits the capabilities of the UV line beam systems. Figure \ref{fig:lineBeamMeasurement} (a) shows a measurements of the $\unit[750]{mm}$ system which has been already introduced in Ref.~\citenum{10.1117/12.2513763} in detail. It consists of two TruMicro 8000 lasers with $\unit[400]{W}$ average power, a beam guiding unit and a line beam optics. The short axis profile can be modified to either Gaussian or flat-top (b, c), beam-widths vary between $\unit[30]{\upmu m}$ and $\unit[45]{\upmu m}$ (FWHM). Additionally, the line length can be decreased which results in possible energy densities of $\unit[\left(300\dots600\right)]{mJ/cm^2}$. Overall, the system combines superior mechanical stability with a very small footprint and is well suited for laser lift-off applications such as debonding of flexible displays. Figure \ref{fig:lineBeamMeasurement} (d) shows an exemplary measurement of a UV-line beam profile with a length of $\unit[1350]{mm}$ and a short axis width of $\unit[90]{\upmu m}$ ($FW$ at $90\,\%$). The long axis homogeneity is in the range of $2\sigma_{\text{LA}} < 1.5\%$ whereas the flat-top profile in the short axis is in the range of $2\sigma_{\text{SA}} < 2.5 \%$. System specifications such as energy density, homogeneity of the intensity distribution and energy stability are designed to meet the requirements for solid state laser annealing of amorphous silicon (a-Si). 
\par
Aside from spatial beam shaping, temporal pulse shaping is of equal significance depending on specific applications. Our beam sources emit pulses in the range of $\tau \sim \unit[20]{ns}$ with up to $\unit[10]{kHz}$ repetition rate. Synchronizing an arbitrary number of lasers and adding defined delays can be used to modify the temporal behaviour of a particular process such as solid state laser annealing of amorphous silicon \cite{10.1117/12.2608238, temporalPulse}. Individual time delays of up to $\unit[500]{ns}$ with a jitter to master clock of $< \unit[2]{ns}$ (1$\sigma$) are possible. As an example, we compare different pulse shapes in Fig.~\ref{fig:lineBeamMeasurement} (f) which can be dynamical created by setting individual time delays between lasers. 

\subsection{Meter-scaled materials processing}\label{sec:multispot}
\begin{figure}[]
    \centering
    \includegraphics[width=0.6\textwidth]{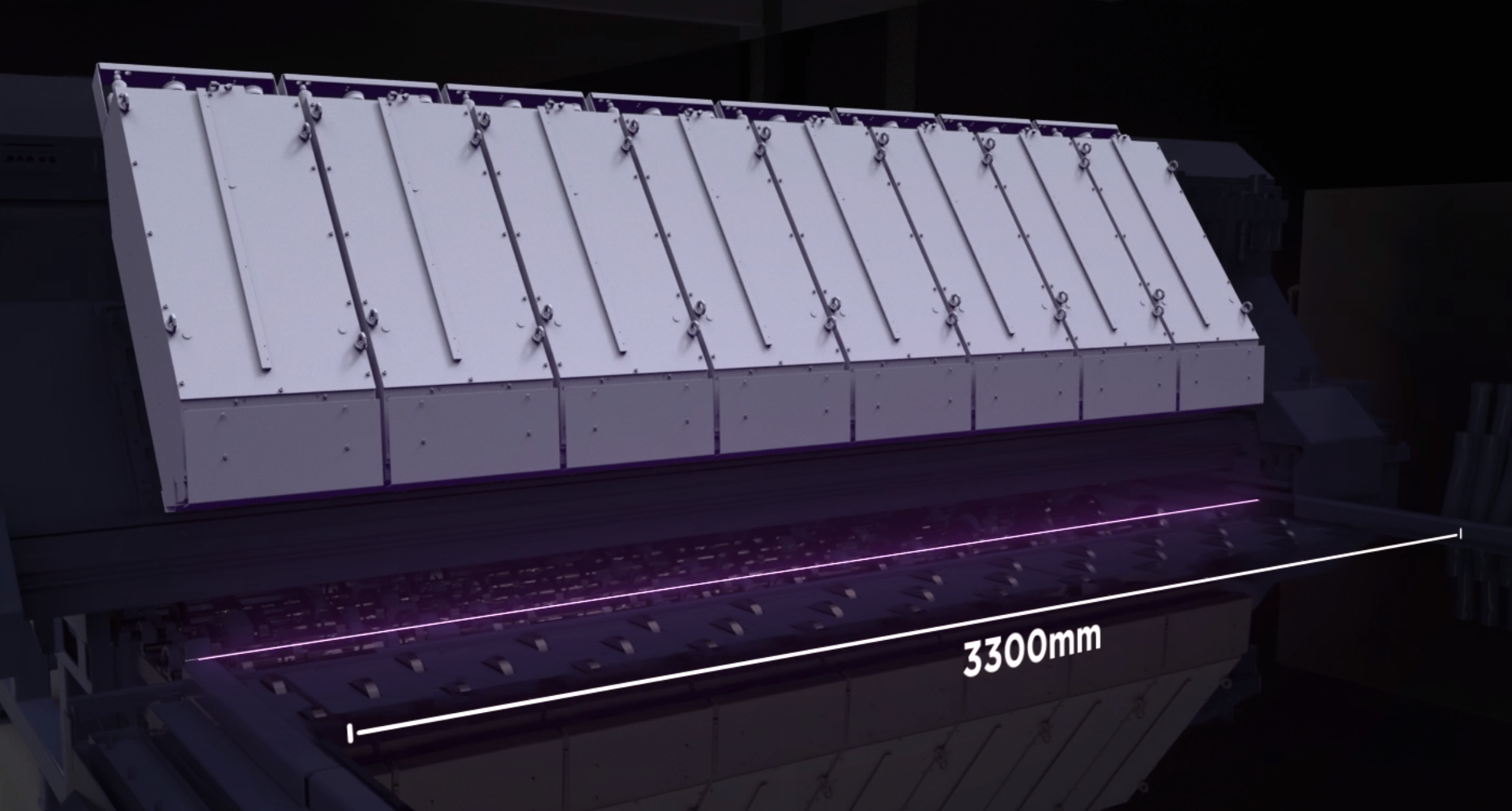}
    \caption{Modular line beam concept for rapid thermal annealing of architectural glass.\cite{IRLinie} The modular design enables the seamless integration of eight line beam modules, resulting in a total laser line length of $\unit[3300]{mm}$. Each individual line beam module generates a $\unit[400]{mm}$ line segment, utilizing the power of a $\unit[18]{kW}$ from TruDisk lasers. Consequently, the combined laser power reaches $\unit[144]{kW}$ ($\unit[40]{W/mm}$).\cite{IRLinie,tillkorn2017novel}}
    \label{fig:IRLine}
\end{figure}

One of the main applications for the line beam system is thermal annealing of, for example, jumbo-size architectural glass.\cite{tillkorn2017novel} Here, highest uniformities are required for focus dimensions beyond $\unit[3]{m}$ to meet the optical demands for coated substrates. At the same time a most confined energy deposition is needed along the short axis of $\sim \unit[60]{\upmu m}$ in order to heat the coating but prevent the substrate to be altered. It has already been demonstrated that IR laser-based annealing decreases the resistivity of, e.g., thin Ag-based stacks due to an improved crystallization process.\cite{tillkorn2017novel} For this application a modular line beam concept is used allowing to seamlessly stitch the line foci generated from eight fiber-coupled modules. Each individual line foci exhibits a length of $\unit[400]{mm}$ utilizing an average power of $\unit[18]{kW}$ from TruDisk lasers, see Fig.~\ref{fig:IRLine}. Consequently, the laser power used in total amounts to $\unit[144]{kW}$ ($\unit[40]{W/mm}$).\cite{IRLinie}
\par 
To outline a second application, we consider the UV line beams system (Sec.~\ref{sec:LBS}) and additionally apply a structured light concept for beam splitting. This creates another subtle laser tool for micromachining on macroscopic scales relevant for display industry. MicroLED represents a promising technology for next generation displays, that moves from niche to mass products \cite{uLED, uLEDTransfer}. However, to accomplish an economic process chain, a transfer approach with a scalable throughput is inevitable---a process efficiently enabled by UV lasers and known as laser-induced forward transfer (LIFT).
Our nanosecond laser source with a repetition rate in the kHz range, combined with beam shaping methods to produce a structured line beam, provides a solution for an economical transfer process.
\par
\begin{figure}[]
    \centering
    \includegraphics[width=0.8\textwidth]{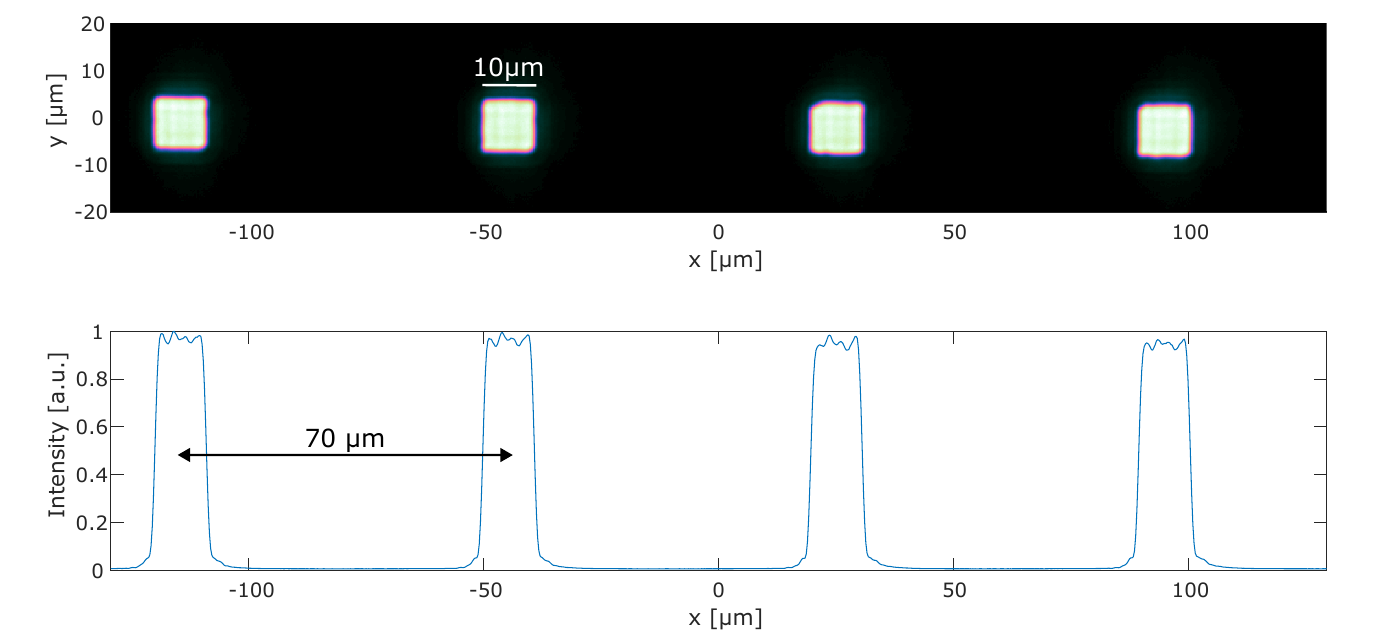}
    \caption{Measurement of a structured line beam profile with homogeneous square shaped flat-tops (top). Cross section of line beam profile, $\unit[10 \times 10]{\mu m^2}$ flat-tops with $\unit[1.5]{\upmu m}$ edge steepness, exemplary pitch of $\unit[70]{\upmu m}$. Homogeneity of plateaus is in the range of $\unit[3.5]{\%}$ ($2\sigma$ at $96\,\%$), which is well suited for successful LIFT (bottom).}
    \label{fig:uLED}
\end{figure}

Enhancing our line beam system to match the requirements for LIFT, high-power capable beam splitting techniques and a diffraction limited imaging system with high numerical aperture ($\text{NA}\sim0.3$) are required. Typically, the ever decreasing micron sized dies are packed much denser on the donor wafer than on the receiver respectively the display. This means, that display resolution and die size results in a specific pitch which µLEDs are needed to be transferred. To avoid illuminating neighboring dies on the densely packed donor wafer, a beam profile with very sharp edges is inevitable. In addition, placement of the µLEDs on the receiver is a key parameter. To prevent µLEDs from misplacement, tilt or flipping, a uniform flat-top profile is necessary to guarantee a homogeneous ablation of the release layer. Figure \ref{fig:uLED} shows a measured intensity profile of a structured line beam where multiple $\unit[\left(10 \times 10\right)]{\upmu m^2}$ square shaped flat-top profiles, separated by an exemplary pitch of $\unit[70]{\upmu m}$ are depicted.
Key parameters such as edge steepness of $\sim \unit[1.5]{\upmu m}$ (steepness $80/20$) and homogeneity of the individual plateaus of roughly $\sim 3.5\,\%$ ($2\sigma$ at $96\,\%$) have been proven to be sufficient for successful LIFT. Therefore our spatially modulated line beam, combined with a multimode beam source and a high-NA diffraction limited imaging system covers all the requirements to accomplish a scalable mass transfer.

\section{CONCLUSION}
\label{sec:con}  
In this work we have reviewed a menu of optical tools enabling laser machining of extreme radiation on extreme spatial dimensions. In any of the cases discussed each optical solution was designed to exploit the entire power and energy performance of the laser system ranging from ultrashort, to short, to continuous wave lasers. Covering even more than six spatial orders of magnitude we introduced the focusing units required and applied structured light concepts as the main enabler for the respective laser application. Optical concepts such as high-NA microscope objectives, large-scanning field F-Theta lenses or line beam optics are designed to handle extreme radiation in terms of average powers, peak powers or wavelengths. An optimum process solution is achieved precisely when light source, beam shaping and focusing concept are tailored to each other and to the respective light-material interaction.

\bibliographystyle{spiebib}   
\bibliography{Lit}

\end{document}